\documentclass[12pt]{article}
\usepackage{amsmath}
\usepackage{graphicx,psfrag,epsf}
\usepackage{enumerate}
\usepackage{natbib}
\usepackage{url} 
\usepackage{mwe}
\usepackage{subfig}
\usepackage{rotating}

\usepackage{amssymb}
\usepackage{amsthm}

\usepackage{amsmath}
\usepackage{multirow}
\usepackage{subfloat}

\newcommand{\blind}{0}

\begin{document}

\def\spacingset#1{\renewcommand{\baselinestretch}%
{#1}\small\normalsize} \spacingset{1}


\if0\blind
{
  \title{\bf Simple Measures of Individual Cluster-Membership Certainty for Hard Partitional Clustering}
  \author{Dongmeng Liu\\
    Statistics and Actuarial Science, Simon Fraser University\\
    and \\
    Jinko Graham\thanks{
    This work was partially supported by \textit{the Natural Sciences and Engineering Research Council of Canada}. The authors thank Joanna Lubieniecka for helpful discussions about clinical databases.}\hspace{.2cm}\\
    Statistics and Actuarial Science, Simon Fraser University}
  \maketitle
} \fi

\if1\blind
{
  \bigskip
  \bigskip
  \bigskip
  \begin{center}
    {\LARGE\bf Simple Measures of Individual Cluster-Membership Certainty for Hard Partitional Clustering}
\end{center}
  \medskip
} \fi

\bigskip
\begin{abstract}
We propose two probability-like measures of individual cluster-membership certainty which can be applied to a hard partition of the sample such as that obtained from the Partitioning Around Medoids (PAM) algorithm, hierarchical clustering or k-means clustering. One measure extends the individual silhouette widths and the other is obtained directly from the pairwise dissimilarities in the sample. Unlike the classic silhouette, however, the measures behave like probabilities and can be used to investigate an individual's tendency to belong to a cluster. We also suggest two possible ways to evaluate the hard partition. We evaluate the performance of both measures in individuals with ambiguous cluster membership, using simulated binary datasets that have been partitioned by the PAM algorithm or continuous datasets that have been partitioned by hierarchical clustering and k-means clustering. For comparison, we also present results from soft clustering algorithms such as soft analysis clustering (FANNY) and two model-based clustering methods. Our proposed measures perform comparably to the posterior-probability estimators from either FANNY or the model-based clustering methods. We also illustrate the proposed measures by applying them to Fisher's classic iris data set.
\end{abstract}

\noindent%
\vfill

\newpage
\spacingset{1.45} 

\section{Introduction}

Clustering is a frequently used method for exploring data. For example, in a clinical study, we may wish to use patient symptoms at diagnosis to identify groups which respond differently to treatment.
One approach to clustering is Bayesian profile regression (\citealp{Molitor2010}), which has the ability to incorporate information on an outcome variable. The profile regression model is fitted to the data by use of a Markov Chain Monte Carlo algorithm, in which the number of clusters and cluster membership changes at each sweep (\citealp{Liverani2015}), and the co-occurrence of a pair of individuals in the same cluster is tracked. After completion of all sweeps, a similarity matrix is created by averaging the pairwise co-occurrences across the sweeps. Then individuals are assigned to clusters by applying the Partitioning Around Medoids or PAM algorithm (\citealp{Kaufman1990}) to the resulting dissimilarity matrix.

One limitation of this approach is that so-called ``hard'' partitional clustering algorithms such as PAM, hierarchical clustering and k-means clustering, assign individuals to distinct clusters but do not provide a measure of the cluster-membership certainties for each individual. Yet, in many applied settings, cluster-membership certainties are desired to help identify individuals with ambiguous group memberships. In hard clustering, one measure of how well an individual belongs to its assigned cluster is the silhouette (\citealp{Rousseeuw1987}). Silhouette values range between negative and positive one, with high values indicating that the individual is well matched to its assigned cluster relative to neighboring clusters. In this note, we propose a simple extension of the silhouette from a single value pertaining to the individual's assigned cluster to a vector of values pertaining to all the clusters in the partition. An attractive feature of the extension is that an individual's values add to one across the clusters and thus provide a probability-like interpretation. Such an interpretation is helpful for assessing the individual's membership uncertainty after the hard clustering has been performed. We also propose another probability-like measure of cluster-membership based directly on the dissimilarity matrix and the partition. The performance of the proposed measures is evaluated in a series of simulation studies. While  model-based and fuzzy-clustering methods give posterior probabilities or so-called memberships to indicate the degree to which individuals belong to each cluster, they are not as commonly used as hard-clustering methods in data exploration. Our proposed measures offer a straightforward way to augment existing output and obtain probability-like measures of cluster-membership certainty for researchers exploring their data with a hard-clustering algorithm. If one wants both a hard and soft classification, our proposed measures are an easy way to obtain that. The measures are computationally quick to calculate, and so soft-membership certainties are conveniently obtained from the hard partition. In the simulation studies, both proposed measures behave similarly to posterior probabilities from model-based and fuzzy clustering. Although our motivation is an application from Bayesian profile regression, the measures can be applied to any pairwise dissimilarity matrix and cluster-membership assignment obtained from hard clustering.

\section{Proposed Measures}
\subsection{Silhouette-Based}
The silhouette is a  widely-used interpretation of how well each individual lies within its assigned cluster (\citealp{Rousseeuw1987}). Each individual's silhouette value is defined by comparing the individual's average dissimilarity with others in its assigned cluster to its dissimilarity with individuals in all other clusters. Let $a_i$ denote the average dissimilarity of individual $i$ with all other individuals within the same cluster and $b_i$ denote the lowest average dissimilarity of individual $i$ to any other cluster, of which $i$ is not a member. The silhouette width is defined as
\begin{equation*}
sil_i = \frac{b_i-a_i}{max\left \{a_i,b_i\right \}}.
\end{equation*}
The range of $sil_i$ is $[-1,1]$. A $sil_i$ close to one indicates that the individual $i$ is appropriately clustered, a $sil_i$ near zero suggests that it lies on the border of two neighboring clusters, and a $sil_i$ close to negative one suggests that it is more appropriately assigned to its neighboring cluster. We extend an individual's silhouette value to a vector, as follows. Given the hard partition, we re-assign the individual to a different cluster holding fixed the other individuals' assignments and compute the corresponding vector of silhouette values for the individual of interest. Since the silhouette values range between -1 and 1, a simple way to make all values positive is to add 1 to every element of the vector. We then add a user-specified exponent, $l$, to the shifted silhouette values and convert them into probabilities by dividing each element by the sum of all elements. 

Let $Z=(z_1, \ldots, z_N)$ denote the cluster-membership assignment or partition for the $N$ individuals in the sample and $C$ denote the number of clusters in the partition. For each individual $i$, we set $z_i=k$ for $k$ in $1 \ldots C$, but leave all remaining elements of $Z$ (for the other individuals) unchanged. Let $sil_{ik} \in [-1,1]$ denote the silhouette value of individual $i$ when the individual is assigned to cluster $k$. Therefore, each individual $i$ is assigned a vector of silhouette values as $(sil_{i1},\ldots,sil_{iC})$. Let $P_{ik}^{(1)}$ denote the silhouette-based measure of cluster-membership certainty for individual $i$ belonging to cluster $k$. Then we define $P_{ik}^{(1)}$ as
\begin{equation} \label{eq:Ph1.1}
P_{ik}^{(1)} = \frac{(sil_{ik}+1)^l}{\sum_{j=1}^C (sil_{ij}+1)^l},
\end{equation}
where $l$ is a user-specified parameter. We call $(sil_{i1}+1, \ldots, sil_{iC}+1)$ the shifted silhouette vector. Each component of the shifted silhouette vector is in the range $[0,2]$. To understand the impact of the exponent term, the ordering of the shifted silhouette values $sil_{i1}+1,\ldots,sil_{iC}+1$ is important. For individual $i$, suppose $sil_{i1}+1>\ldots>sil_{iC}+1$. Increasing the exponent term $l$ pushes the measure $P_{i1}^{(1)}$ closer to 1 because $(sil_{i1}+1)^l$ increases relative to $(sil_{ik}+1)^l$ for $k=2,\ldots,C$, when $l$ increases. Large values of $l$ should therefore produce crisper clusters.

\subsection{Dissimilarity-Based}
In addition to the silhouette-based measure, we propose a measure that is based directly on the pairwise-dissimilarity matrix. Assume that the pairwise-dissimilarity matrix, $\{d_{ij} \}$, between $N$ individuals is given, and has non-negative entries. Let $h_{ik} > 0$ be the average dissimilarity between individual $i$ and cluster $k$ such that
\begin{equation*}
h_{ik} =   \left ( \sum_{ j  \not= i: Z_j =k} d_{ij}  \right )  \; \bigg/ \;
| \{ j \not= i: Z_j =k \} |,
\end{equation*}
where $| \{ j \not= i: Z_j =k \} |$ denotes the number of all other individuals in cluster $k$ except individual $i$. As $h_{ik}$ is an overall measure of dissimilarity between individual $i$ and cluster $k$, we may consider $s_{ik}=1/h_{ik}>0$ as an overall measure of similarity. The higher the $s_{ik}$, the better individual $i$ fits into cluster $k$. A measure of cluster-membership certainty of individual $i$ belonging to cluster $k$ is therefore
\begin{equation} \label{eq:Ph1.2}
P_{ik}^{(2)} =\frac{s_{ik}^v}{\sum_{j=1}^C s_{ij}^v},
\end{equation}
where $v$ is a user-specified exponent. To understand the impact of the exponent term, the ordering of the similarities $s_{i1},\ldots,s_{iC}$ is important. For individual $i$, suppose $s_{i1}>\ldots>s_{iC}$. Increasing the exponent term $v$ pushes the measure $P_{i1}^{(2)}$ closer to one, since $s_{i1}^v$ increases relative to $s_{ik}^v$ for $k=2,\ldots,C$, when $v$ increases. As a result, large values of $v$ lead to crisper clusters.

\section{Evaluation For A Hard Partition}

We suggest two ways to use the proposed measures to evaluate the clustering solution obtained from a hard partition, via the soft-misclassification rate and the partition-disagreement rate. Let the soft-misclassification rate be $R_{sm}^{(q)}$, where $q=1,2$ for the silhouette- and dissimilarity-based measure, respectively. Let $g_i$ denote the true group of individual $i$; then the soft-misclassification rate is defined as
\begin{equation} \label{eq:softmis}
R_{sm}^{(q)}=\frac{1}{N} \sum_{i=1}^N (1-P_{ig_i}^{(q)}),
\end{equation}
The soft-misclassification rate weights crisp cluster memberships differently than fuzzy memberships. Specifically, the higher the membership certainty for the true group of an individual, the lower the contribution of that individual to the soft-misclassification rate. 

One drawback of the soft-misclassification rate is that we require the true assignment which may not be available in practice. When the true assignment is unknown, we suggest a partition-disagreement rate, which addresses the disagreement between the hard partition and the individual measures of cluster-membership certainty. Recall that $z_i$ denotes the assigned cluster of individual $i$; then the disagreement rate is 
\begin{equation} \label{eq:disagreement}
R_{pd}^{(q)}=\frac{1}{N} \sum_{i=1}^N (1-P_{iz_i}^{(q)}), 
\end{equation}
where $q=1,2$ for the silhouette-based and the dissimilarity-based measure, respectively. A large value of the partition-disagreement rate suggests a potential disagreement between the hard partition and the dissimilarity matrix. 

When the hard partition is consistent with the true assignment, the soft-misclassification rate and the partition-disagreement rate are equal.

\section{Simulation Study}
In this section, we consider and simulate four situations: two groups of individuals with binary features, three groups with binary features, two groups with continuous features and three groups with continuous features. In each situation, the groups are easily differentiated by the clustering methods and an individual is added as a hybrid of the groups. Figure 1a-d shows a typical data structure for each simulated situation.

\begin{figure}
\begin{minipage}{.05\linewidth}
\end{minipage}
\begin{minipage}{.47\linewidth}
\centering
\text{two-group}
\end{minipage}
\begin{minipage}{.47\linewidth}
\centering
\text{three-group}
\end{minipage}
\begin{minipage}{.05\linewidth}
\rotatebox{90}{binary}
\end{minipage}
\begin{minipage}{.47\linewidth}
\centering
\subfloat[]{\label{fig:sub1}\includegraphics[height=3in, width=3in]{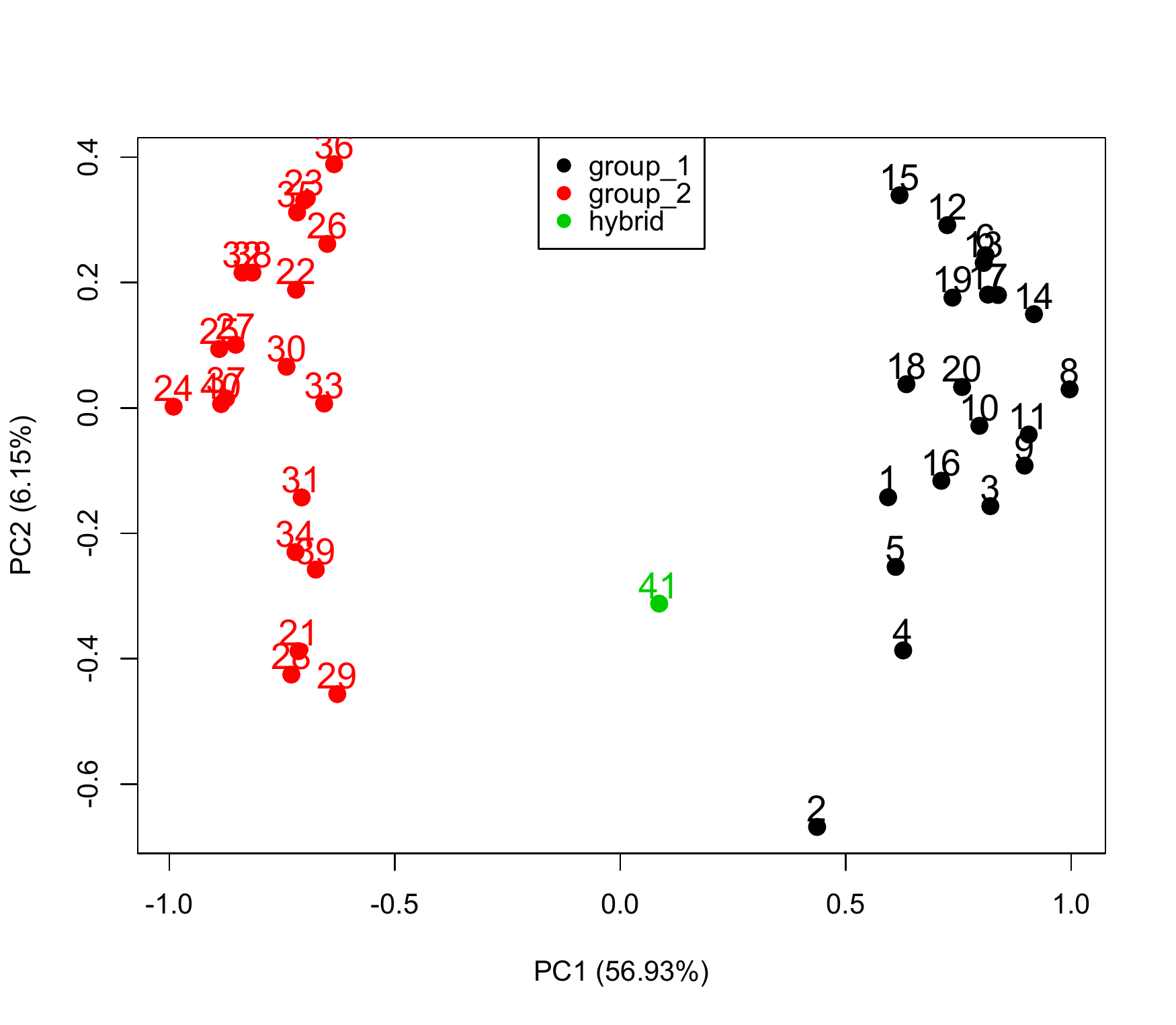}}
\end{minipage}
\begin{minipage}{.47\linewidth}
\centering
\subfloat[]{\label{fig:sub2}\includegraphics[height=3in, width=3in]{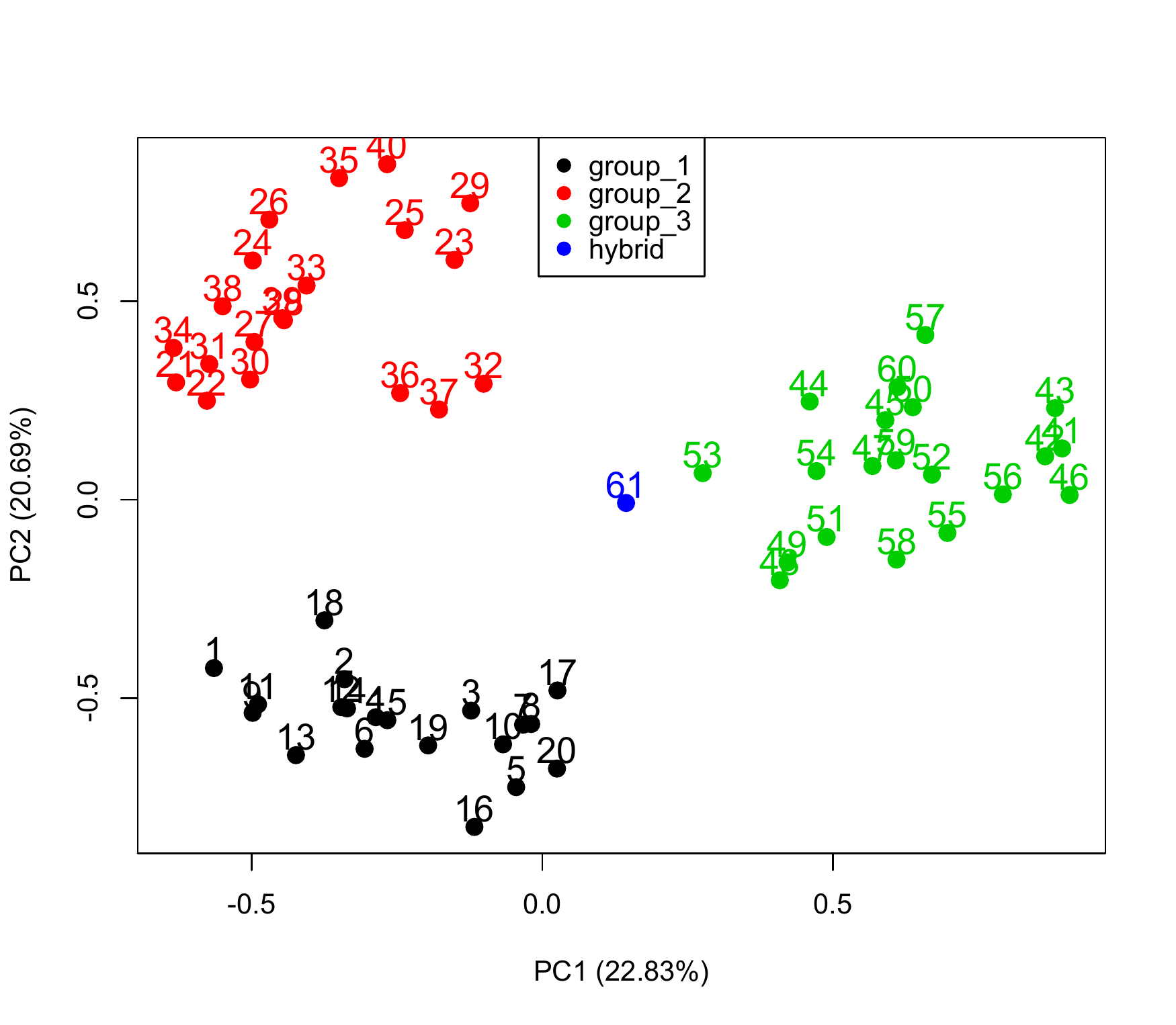}}
\end{minipage}
\begin{minipage}{.05\linewidth}
\rotatebox{90}{continuous}
\end{minipage}
\begin{minipage}{.47\linewidth}
\centering
\subfloat[]{\label{fig:sub3}\includegraphics[height=3in, width=3in]{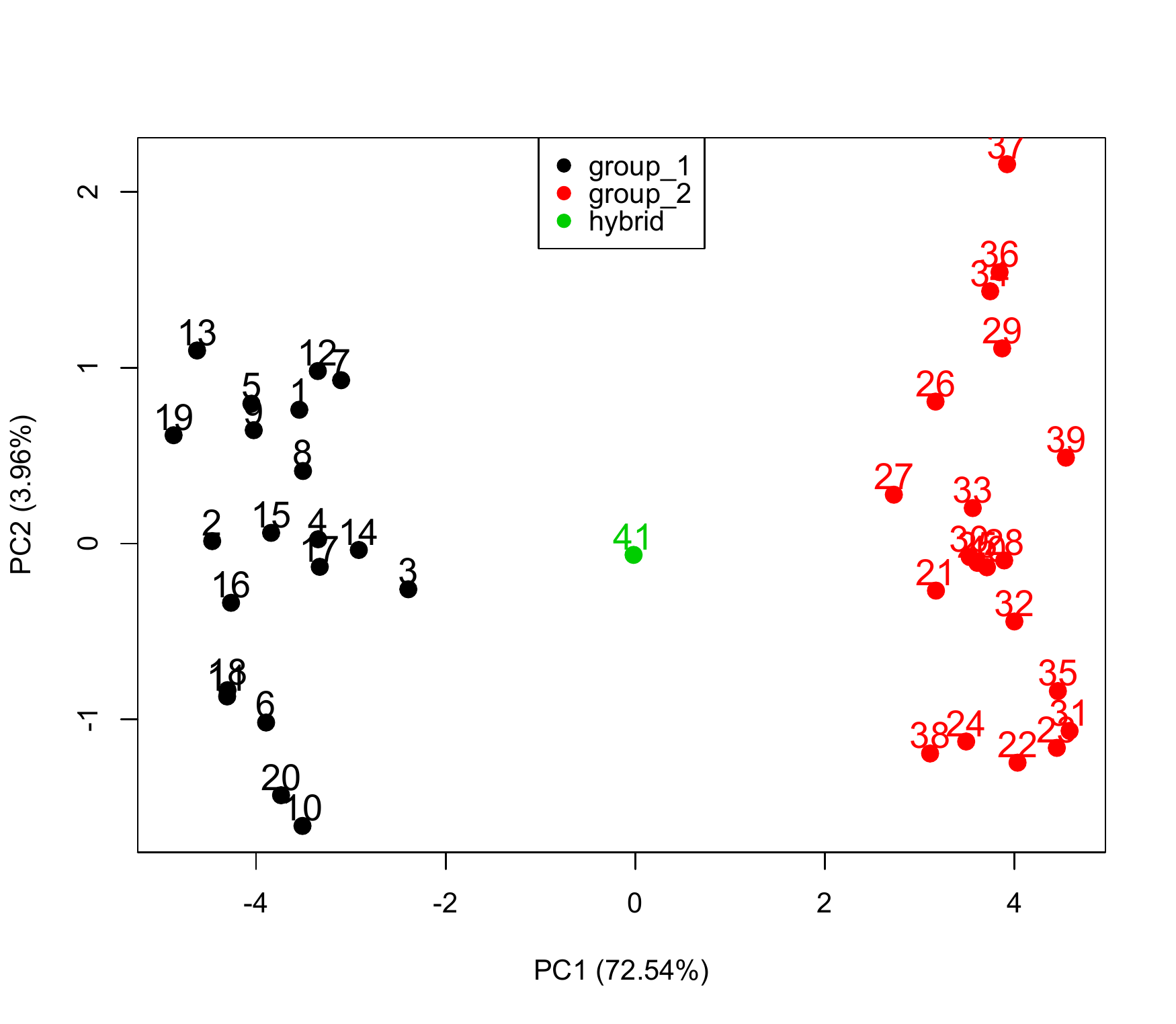}}
\end{minipage}
\begin{minipage}{.47\linewidth}
\centering
\subfloat[]{\label{fig:sub4}\includegraphics[height=3in, width=3in]{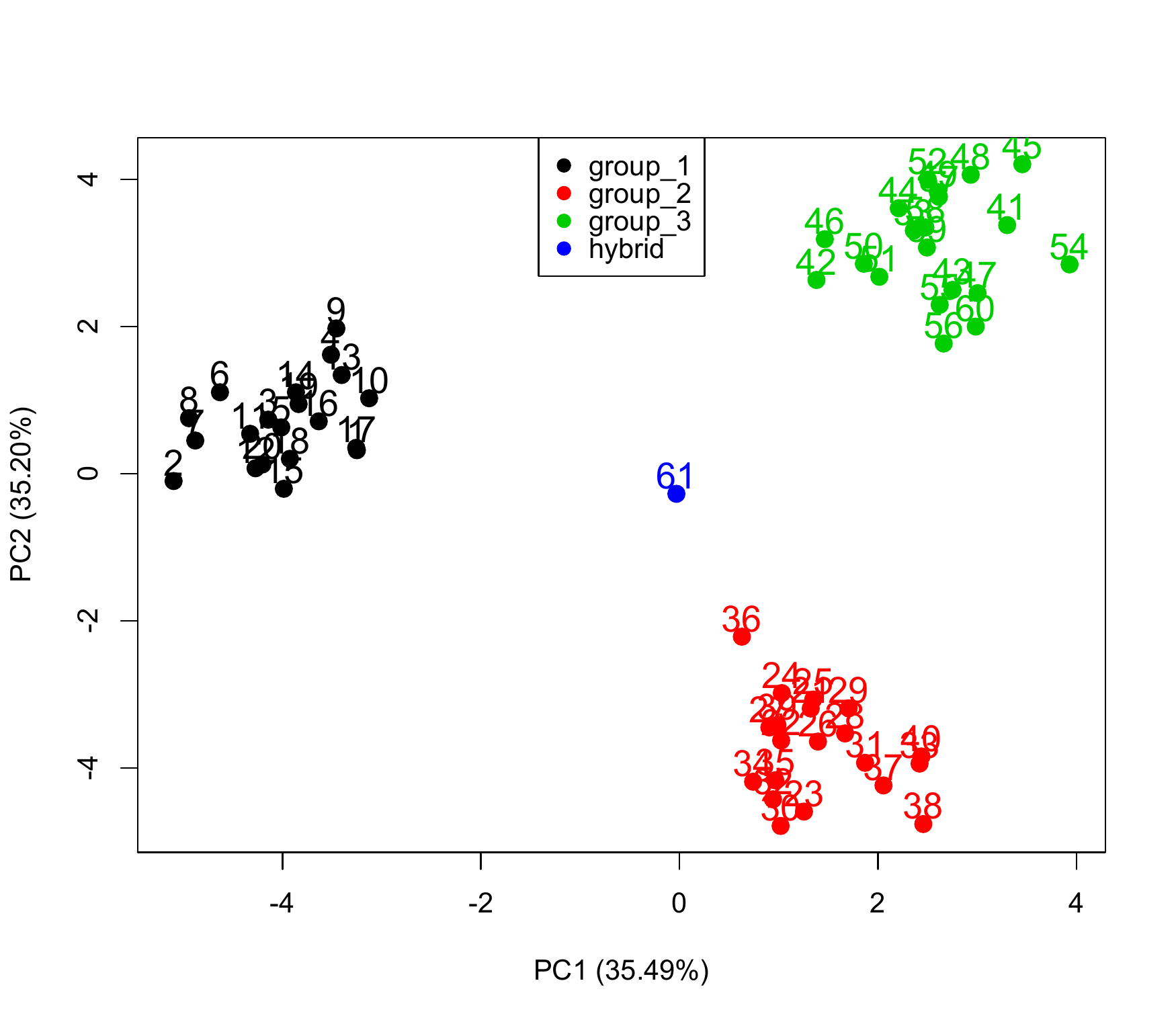}}
\end{minipage}
\caption{Typical data structure for simulated datasets. (a) A plot of the first two principal coordinates from a multiple correspondence analysis for the two-group binary datasets. (b) The first two principal coordinates for the three-group binary datasets. (c) The first two principal components from PCA for two-group continuous datasets. (d) The first two principal components for the three-group continuous datasets.}
\label{fig:datastructure}
\end{figure}

\subsection{Binary Data}

We consider a number of possible dissimilarity matrices in our simulations of the binary data: Euclidean distance based on the top two principal coordinates from multiple correspondence analysis, simple matching distance (SMD) (see, e.g. \citealp{Gower2004}) and the {\tt PReMiuM} co-occurrence dissimilarity from profile regression described in the introduction. We evaluate the proposed measures of cluster-membership certainty using the hard partition assigned by PAM. However, a partition from hierarchical clustering or from k-means clustering applied to the principal coordinates from MCA could also be used. As a benchmark for comparison, we also apply soft-clustering and compute the posterior probability of belonging to cluster 1 under (i) LCA applied directly to the discrete data on the features (see, e.g. \citealp{Lazarsfeld1968} and \citealp{McCutcheon1987}), (ii) a Gaussian mixture model (\citealp{Banfield1991}) applied to the top two principal coordinates obtained from multiple correspondence analysis and (iii) the FANNY algorithm's memberships (\citealp{Kaufman1990}).

\subsubsection{Two Groups}

We first evaluate the cluster-membership certainty of an individual which is a hybrid of two groups. Group membership is determined by two latent variables, $U^{(1)}$ and $U^{(2)}$,  each taking on two values.
From each of the two groups, 20 individuals are simulated with 20 binary features. We assign individuals $1, \ldots, 20$ to group 1 and individuals $21, \ldots, 40$ to group 2. Let $X_{ij}\in\{0,1\}$ denote the $j$th feature of individual $i$, and $U_i^{(1)}$ and $U_i^{(2)}$ the latent variables for individual $i$. 
The conditional probability of a binary feature being a success is
\begin{eqnarray}
Pr(X_{ij}=1 \mid U_i^{(1)}, U_i^{(2)}) =
\begin{cases}
\frac{exp(t+\beta*U_i^{(1)})}{1+exp(t+\beta*U_i^{(1)})} & \text{for } j = 1,\ldots,10\\
\frac{exp(t+\beta*U_i^{(2)})}{1+exp(t+\beta*U_i^{(2)})} & \text{for } j = 11,\ldots,20,
\end{cases}
\label{eq:binaryfeature}
\end{eqnarray}
where the intercept $t$ is selected to ensure that 
\begin{equation*}
Pr(X=1) = \sum_{u^{(1)}, u^{(2)}} Pr(X=1 \mid U^{(1)}=u^{(1)}, U^{(2)}=u^{(2)}) Pr(U^{(1)}=u^{(1)}, U^{(2)}=u^{(2)})=0.5.
\end{equation*}

We set $\beta=1.2$;  $(U_i^{(1)},U_i^{(2)})=(3,0)$ for group 1, $i=1,\ldots,20$; $(U_i^{(1)},U_i^{(2)})=(0,3)$ for group 2, $i=21,\ldots,40$; and $(U_i^{(1)},U_i^{(2)})=(1.5,1.5)$ for the hybrid individual, $i=41$. These values ensure that the two groups can be easily differentiated on a plot of the first two principal coordinates from a multiple correspondence analysis (see \citealp{Le2004}). Figure \ref{fig:sub1} shows a typical structure of the simulated data set, with group 1 and 2 represented by black and red points, and the hybrid individual labelled as 41 in green. For the 40 non-hybrid individuals, we define clusters 1 and 2 as those assigned more individuals coming from true groups 1 and 2, respectively. An individual is classified correctly if the indices of its true group and cluster assignment agree, and misclassified otherwise. We compute $P_{h1}^{(1)}$ and $P_{h1}^{(2)}$ as in equations \eqref{eq:Ph1.1} and \eqref{eq:Ph1.2}; i.e., as the cluster-membership certainties of the hybrid individual for cluster 1 when the number of clusters is fixed to 2. 

We simulate 1000 datasets to obtain the empirical distribution of the hybrid's cluster-membership certainties. Since hybrid individuals are equally distant from either group, we expect the distributions of $P_{h1}^{(1)}$ and $P_{h1}^{(2)}$ to be symmetric, with a mean around 0.50. We also consider the 40 non-hybrid individuals and calculate their soft-misclassification rate $R_{sm}$ and partition-disagreement rate $R_{pd}$ as in equations \eqref{eq:softmis} and \eqref{eq:disagreement}.  

\subsubsection{Three Groups}

In addition to the simulation studies with two groups, we also simulate data with three groups and a hybrid individual. Group membership is determined by three latent variables, $U^{(1)}$, $U^{(2)}$ and $U^{(3)}$. We assign individuals 1,$\ldots$,20 to group 1, individuals 21, $\ldots$,40 to group 2 and individuals 41,$\ldots$,60 to group 3. Each individual is simulated with 24 binary features, with features $X_1,\ldots,X_{8}$ being determined by $U^{(1)}$, features $X_9,\ldots,X_{16}$ by $U^{(2)}$ and features $X_{17},\ldots,X_{24}$ by $U^{(3)}$. The conditional probability of a binary feature being a success is computed similarly to equation \eqref{eq:binaryfeature}. Briefly, we set $\beta=1.2$;  $(U_i^{(1)},U_i^{(2)}, U_i^{(3)})=(3,0,0)$ for group 1, $i=1,\ldots,20$; $(U_i^{(1)},U_i^{(2)}, U_i^{(3)})=(0,3,0)$ for group 2, $i=21,\ldots,40$; $(U_i^{(1)},U_i^{(2)}, U_i^{(3)})=(0,0,3)$ for group 3, $i=41,\ldots,60$; and $(U_i^{(1)},U_i^{(2)}, U_i^{(3)})=(1,1,1)$ for the hybrid individual, $i=61$. A typical structure of the simulated data set based on the first two principal coordinates obtained from MCA is shown in Figure \ref{fig:sub2}. The hybrid individual is labelled as 61. 

Similar to the two-group setting, we simulate 1000 datasets and compute the cluster-membership certainties of the hybrid individual for cluster 1 when the number of clusters is fixed to 3. For the hybrid individual, we expect the distribution of $P_{h1}^{(1)}$ and $P_{h1}^{(2)}$ to center around 0.33. The soft-misclassification rate and partition-disagreement rate of the 60 non-hybrid individuals are also computed.



\subsection{Continuous Data}

Similar to the binary datasets, we simulate two or three well-separated groups with a hybrid individual. To measure the dissimilarity between two individuals, we use the Euclidean distance between their feature vectors. We then apply hierarchical clustering and k-means clustering methods, and compute the cluster-membership certainty for the hybrid individual. We use the Gaussian mixture model and the FANNY algorithm as benchmarks since LCA is for binary. For each setting, we simulate 1000 datasets and expect the distribution of $P_{h1}$ for the hybrid to center around 0.50 for the two-group datasets and 0.33 for the three-group datasets.

\subsubsection{Two Groups}

For the two-group datasets, group membership is determined by two latent variables, $W^{(1)}$ and $W^{(2)}$. For each group, we simulate 20 individuals with 20 features. The features are simulated as $X_{ij} \sim N(W_i^{(1)},1)$, for $j=1,\ldots,10$ and $X_{ij} \sim N(W_i^{(2)},1)$, for $j=11,\ldots,30$, where $i$ indexes individuals and $j$ indexes features. We set $(W_i^{(1)},W_i^{(2)})=(3,0)$ for group 1, $i=1,\ldots,20$; $(W_i^{(1)},W_i^{(2)})=(0,3)$ for group 2, $i=21,\ldots,40$; and $(W_i^{(1)},W_i^{(2)})=(1.5,1.5)$ for the hybrid individual, $i=41$. Referring to Figure \ref{fig:sub3}, group 1 (black) and  group 2 (red) are well-separated with the hybrid individual labeled as 41 in green, placed between.

\subsubsection{Three Groups}

For the three-group datasets, group membership is determined by three latent variables, $W^{(1)}$, $W^{(2)}$ and $W^{(3)}$. For each group, we simulate 20 individuals with 24 features. The features are simulated as $X_{ij} \sim N(W_i^{(1)},1)$ for $j=1,\ldots,8$, $X_{ij} \sim N(W_i^{(2)},1)$ for $j=9,\ldots,16$ and $X_{ij} \sim N(W_i^{(3)},1)$ for $j=17,\ldots,24$. We set $(W_i^{(1)},W_i^{(2)},W_i^{(3)})=(3,0,0)$ for group 1, $i=1,\ldots,20$; $(W_i^{(1)},W_i^{(2)},W_i^{(3)})=(0,3,0)$ for group 2, $i=21,\ldots,40$; $(W_i^{(1)},W_i^{(2)},W_i^{(3)})=(0,0,3)$ for group 3, $i=41,\ldots,60$; and $(W_i^{(1)},W_i^{(2)},W_i^{(3)})=(1,1,1)$ for the hybrid individual, $i=61$. Figure \ref{fig:sub4} shows a typical structure of the dataset based on the first two principal components. Group 1 (black), group 2 (red) and group 3 (green) are well-separated, with the hybrid individual labeled as 61 in blue, placed between.
 
\subsection{Implementation}

The simulation study is implemented in R. The R package {\tt FactoMineR} (\citealp{Le2008}) provides functions for multiple correspondence analysis and data visualization. The profile regression mixture model is implemented in the R package {\tt PReMiuM} (\citealp{Liverani2015}). The PAM and FANNY algorithms are implemented in the R package {\tt cluster}. The hierarchical clustering and k-means clustering are implemented in the R package {\tt stats} (\citealp{Rstats}). The implementation of LCA for binary covariates is available in the R package {\tt poLCA} (\citealp{Linzer2011}; R Core Team, 2012). The Gaussian mixture model is implemented in the R package {\tt mclust} (see \citealp{Fraley2012} and \citealp{Fraley2002}).

\section{Results}

\subsection{Proposed Measures}

\subsubsection{Binary Data}

For both of the proposed measures, the tuning parameter changes the soft-misclassification rate and partition-disagreement rate, and the distributional shape of the hybrid's cluster-membership certainty. Tables \ref{tab:tuning_binary_2clst} and \ref{tab:tuning_binary_3clst} show the effect of tuning the exponent parameter of the two measures for the two-group and three-group datasets, respectively. In these tables, we fix the standard deviations at arbitrary values of 0.15, 0.20 and 0.25 and present the corresponding tuning parameters ($l$ or $v$), soft-misclassification rates ($R_{sm}$) and partition-disagreement rates ($R_{pd}$). The means of $P_{h1}$ are 0.50 and 0.33 for the two-group and three-group datasets, respectively, regardless of the dissimilarity matrix or the value of tuning parameter, as expected for the hybrid individual (results not shown). In general, the exponent parameters $l$ and $v$ represent a tradeoff between detecting the hybrid individual with $P_{h1}$ and minimizing the soft-misclassification rate $R_{sm}$ or partition-disagreement rate $R_{pd}$, for the non-hybrid individuals. Specifically, increasing $l$ and $v$ leads to larger variance in $P_{h1}^{(1)}$ and $P_{h1}^{(2)}$ but lower $R_{sm}$ and $R_{pd}$. For example, referring to the entry of the silhouette-based measure of the Euclidean distance matrix in the first row of Table \ref{tab:tuning_binary_2clst}, increasing the tuning parameter $l$ from 0.9 to 1.8 increases ${\rm sd}(P_{h1}^{(1)})$ from 0.15 to 0.25 while decreasing both $R_{sm}^{(1)}$ and $R_{pd}^{(1)}$ from 14.85\% to 3.47\%.

\begin{figure}
\begin{minipage}{1\linewidth}
\centering
\subfloat[]{\label{fig:hist1}\includegraphics[height=1.7in, width=3.4in]{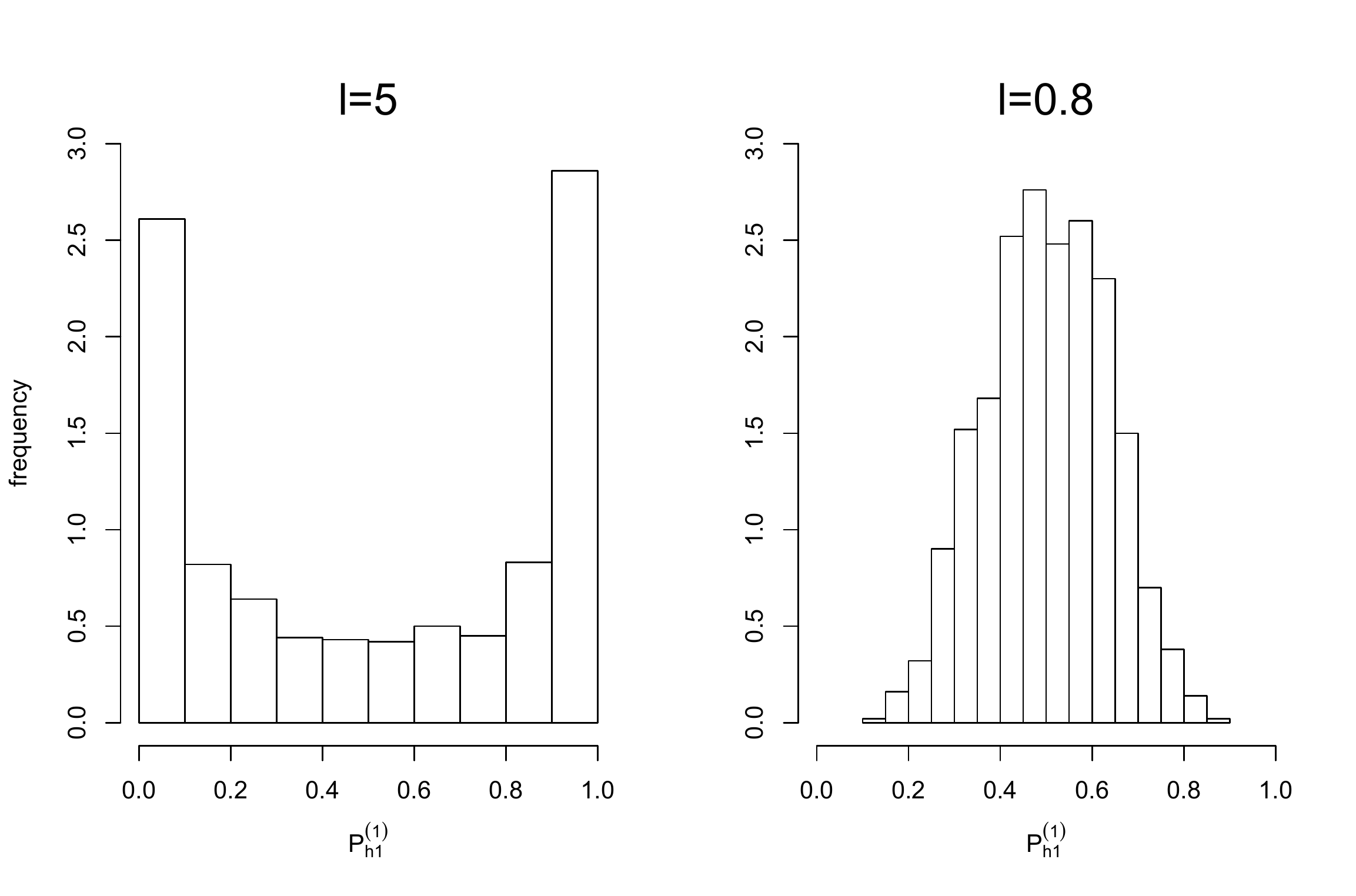}}
\end{minipage}
\begin{minipage}{1\linewidth}
\centering
\subfloat[]{\label{fig:hist2}\includegraphics[height=1.7in, width=3.4in]{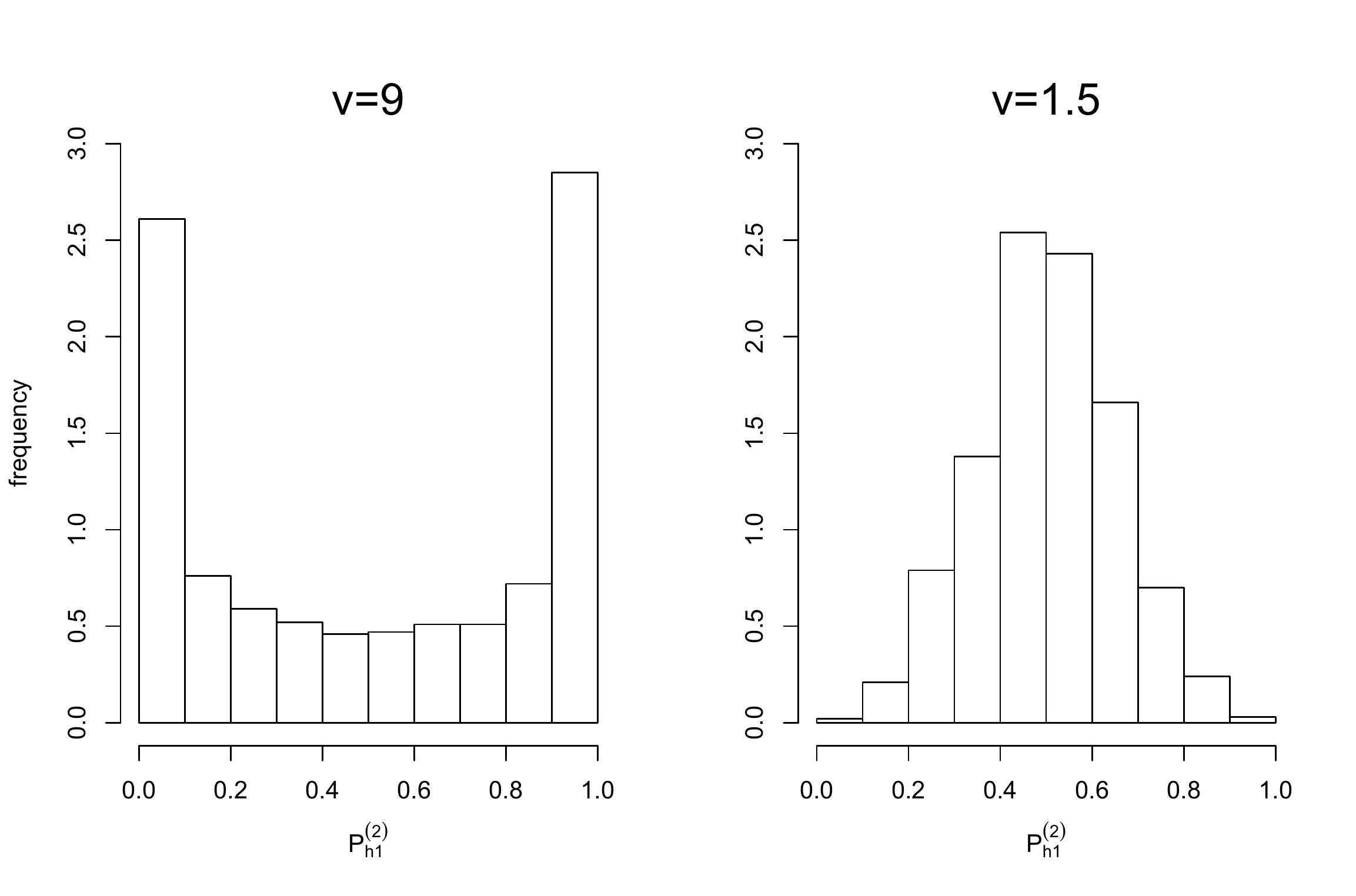}}
\end{minipage}
\caption{In panel (a), decreasing $l$ from 5 to 0.8 changes the respective distribution of $P_{h1}^{(1)}$ from being symmetric with a U-shape to being symmetric with a mode at 0.5. In panel (b), decreasing $v$ from 9 to 1.5 changes the respective distribution of $P_{h1}^{(2)}$ from being symmetric with a U-shape to being symmetric with a mode at 0.5. Both $P_{h1}^{(1)}$ and $P_{h1}^{(2)}$ are shown for the Euclidean distance matrix of the two-group datasets.}
\label{fig:tuning}
\end{figure}

Figure \ref{fig:tuning} shows an example of how tuning $l$ and $v$ influences the shape of the distribution of $P_{h1}^{(1)}$ and $P_{h1}^{(2)}$ based on the Euclidean distance matrix of the two-group datasets. For the silhouette-based measure $P_{h1}^{(1)}$ in panel (a), $P_{h1}^{(1)}$ is more variable with a large $l=5$, where most values are close to 0 or 1; as $l$ decreases to $l=0.8$, $P_{h1}^{(1)}$ takes on less extreme values and still has a mode at 0.50. Similarly, for the dissimilarity-based measure in panel (b),  most of the values of $P_{h1}^{(2)}$ are close to either 0 or 1 when $v=9$ but, when $v=1.5$, they tend to concentrate around 0.50. 

The performance of $P_{h1}$ depends on the dissimilarity matrices. For example, for a fixed value of the tuning parameter, the measures of cluster-membership certainty for the hybrids based on the SMD matrix are more concentrated about the true values than those based on the Euclidean distance or {\tt PReMiuM} dissimilarity matrices. However, for the non-hybrid individuals, the contrast between the within- and between-cluster similarities based on the SMD matrix is not as stark as for the Euclidean distance and {\tt PReMiuM} co-occurrence matrices and so the misclassification/disagreement rates for the SMD matrix are larger (see Table 3 and Section 5.2.1 for an example of the default values of the tuning parameters). Thus, to generate a given standard deviation of the hybrid's cluster-membership certainties, the SMD dissimilarity matrix requires a larger value of the tuning parameter than the Euclidean distance or {\tt PReMiuM} dissimilarity matrices. The larger value of the tuning parameter results in a relatively small value of the soft-misclassification rate and the partition-disagreement rate for the non-hybrids.  

Moving to a comparison of the proposed measures for a given dissimilarity matrix, generally speaking, the silhouette-based measure produces crisper cluster-membership certainties than the dissimilarity-based measure. Thus in our simulations, the silhouette-based measure tends to have smaller values of the tuning parameter ($l$) than the dissimilarity-based measure ($v$) for a targeted standard deviation of the hybrid's cluster-membership certainty. One advantage of tuning parameters is that, for $l$ and $v$ tuned to give the same standard deviation, the two measures generate similar rates of both soft-misclassification and partition-disagreement given a dissimilarity matrix, as shown in Table \ref{tab:tuning_binary_2clst}.

\begin{subtables}
\begin{table}[htbp]
\centering
\begin{tabular}{c|c|ccc|ccc|ccc}
\multirow{2}{*}{\begin{tabular}[c]{@{}c@{}}dissimilarity\\ matrix\end{tabular}} & \multirow{2}{*}{measure} & \multicolumn{3}{c|}{sd=0.15} & \multicolumn{3}{c|}{sd=0.20} & \multicolumn{3}{c}{sd=0.25} \\ \cline{3-11} 
 &  & $R_{sm}$ & $R_{pd}$ & $l$ or $v$ & $R_{sm}$ & $R_{pd}$ & $l$ or $v$ & $R_{sm}$ & $R_{pd}$ & $l$ or $v$ \\ \hline
 \multirow{2}{*}{Euclidean} & silhouette & 14.85 & 14.85 & 0.9 & 7.78 & 7.78 & 1.3 & 3.47 & 3.47 & 1.8 \\
 & dissimilarity & 11.38 & 11.38 & 1.5 & 5.22 & 5.22 & 2.2 & 2.26 & 2.26 & 3.0 \\ \hline
\multirow{2}{*}{SMD} & silhouette & 12.11 & 12.11 & 2.2 & 6.25 & 6.25 & 3.1 & 2.65 & 2.65 & 4.3 \\
 & dissimilarity & 10.66 & 10.66 & 3.9 & 5.13 & 5.13 & 5.6 & 2.11 & 2.11 & 7.8 \\ \hline
\multirow{2}{*}{{\tt PReMiuM}} & silhouette & 14.75 & 14.75 & 0.4 & 7.49 & 7.49 & 0.6 & 2.85 & 2.85 & 0.9 \\
 & dissimilarity & 13.83 & 13.83 & 0.5 & 6.16 & 6.16 & 0.8 & 2.33 & 2.33 & 1.2
\end{tabular}
\caption{For the two-group binary datasets, the soft-misclassification rates $R_{sm}$ and partition-disagreement rates $R_{pd}$ as percents for different values of ${\rm sd}(P_{h1})$ and three dissimilarity matrices. The value of the tuning parameter, $l$ or $v$, used to achieve the ${\rm sd}(P_{h1})$ is also shown} 
\label{tab:tuning_binary_2clst}
\end{table}

\begin{table}[htbp]
\centering
\begin{tabular}{c|c|ccc|ccc|ccc}
\multirow{2}{*}{\begin{tabular}[c]{@{}c@{}}dissimilarity\\ matrix\end{tabular}} & \multirow{2}{*}{measure} & \multicolumn{3}{c|}{sd=0.15} & \multicolumn{3}{c|}{sd=0.20} & \multicolumn{3}{c}{sd=0.25} \\ \cline{3-11} 
 &  & $R_{sm}$ & $R_{pd}$ & $l$ or $v$ & $R_{sm}$ & $R_{pd}$ & $l$ or $v$ & $R_{sm}$ & $R_{pd}$ & $l$ or $v$ \\ \hline
\multirow{2}{*}{Euclidean} & silhouette & 23.77 & 23.75 & 1.0 & 15.65 & 15.65 & 1.3 & 7.64 & 7.64 & 1.8 \\ 
 & dissimilarity & 23.78 & 23.77 & 1.4 & 13.22 & 13.19 & 2.0 & 6.10 & 6.07 & 2.9 \\ \hline
\multirow{2}{*}{SMD} & \multicolumn{1}{c|}{silhouette} & 16.87 & 16.98 & 4.4 & 8.63 & 8.79 & 6.2 & 4.08 & 4.29 & 8.4 \\
 & \multicolumn{1}{c|}{dissimilarity} & 18.04 & 18.14 & 7.2 & 9.15 & 9.29 & 10.4 & 4.42 & 4.61 & 14.2 \\ \hline
\multirow{2}{*}{{\tt PReMiuM}} & \multicolumn{1}{c|}{silhouette} & 22.63 & 22.59 & 0.4 & 14.14 & 14.09 & 0.6 & 5.76 & 5.68 & 0.9 \\
 & dissimilarity & 21.81 & 21.77 & 0.6 & 11.70 & 11.65 & 0.8 & 4.73 & 4.66 & 1.2
\end{tabular}
\caption{For the three-group binary datasets, the soft-misclassification rates $R_{sm}$ and partition-disagreement rates $R_{pd}$ as percents for different values of ${\rm sd}(P_{h1})$ and three dissimilarity matrices. The value of the tuning parameter, $l$ or $v$, used  to achieve the ${\rm sd}(P_{h1})$ is also shown} 
\label{tab:tuning_binary_3clst}
\end{table}
\end{subtables}

\subsubsection{Continuous Data}

For the continuous datasets, the tuning parameter changes the soft-misclassification rate and partition-disagreement rate, as well as the distributional shape of the hybrid's cluster-membership certainty in a similar way to the binary data. Tables \ref{tab:tuning_cont_2clst} and \ref{tab:tuning_cont_3clst} show the effect of tuning the exponent parameters for the two-group datasets and three-group datasets, respectively. In these tables, the standard deviations are fixed at 0.05, 0.10 and 0.15 and the corresponding tuning parameters, soft-misclassification rates and partition-disagreement rates are presented. For both clustering methods, the means of $P_{h1}$ are 0.50 and 0.33 for the two-group and three-group datasets, respectively, regardless of the value of the tuning parameter. Increasing $l$ and $v$ leads to larger variance in $P_{h1}^{(1)}$ and $P_{h1}^{(2)}$ but lower soft-misclassification and partition-disagreement rates.

\begin{subtables}
\begin{table}[htbp]
\centering
\begin{tabular}{c|c|ccc|ccc|ccc}
\multirow{2}{*}{\begin{tabular}[c]{@{}c@{}}clustering\\ method\end{tabular}} & \multirow{2}{*}{measure} & \multicolumn{3}{c|}{sd=0.05} & \multicolumn{3}{c|}{sd=0.10} & \multicolumn{3}{c}{sd=0.15} \\ \cline{3-11} 
 &  & $R_{sm}$ & $R_{pd}$ & $l$ or $v$ & $R_{sm}$ & $R_{pd}$ & $l$ or $v$ & $R_{sm}$ & $R_{pd}$ & $l$ or $v$ \\ \hline
\multirow{2}{*}{hierarchical} & silhouette & 28.83 & 28.83 & 0.7 & 14.18 & 14.86 & 1.4 & 5.66 & 5.66 & 2.2 \\
 & dissimilarity & 25.10 & 25.14 & 1.3 & 10.25 & 10.25 & 2.5 & 3.53 & 3.53 & 4.0 \\ \hline
\multirow{2}{*}{k-means} & silhouette & 28.82 & 28.82 & 0.7 & 14.18 & 15.08 & 1.4 & 5.66 & 5.66 & 2.2 \\
 & dissimilarity & 25.09 & 25.12 & 1.3 & 10.24 & 10.28 & 2.5 & 3.53 & 3.55 &4.0
\end{tabular}
\caption{For the two-group continuous datasets, the soft-misclassification rates $R_{sm}$ and partition-disagreement rates $R_{pd}$ as percents for different values of ${\rm sd}(P_{h1})$ and two different clustering methods. The value of the tuning parameter, $l$ or $v$, used to achieve the ${\rm sd}(P_{h1})$ is also shown} 
\label{tab:tuning_cont_2clst}
\end{table}

\begin{table}[htbp]
\centering
\begin{tabular}{c|c|ccc|ccc|ccc}
\multirow{2}{*}{\begin{tabular}[c]{@{}c@{}}clustering\\ method\end{tabular}} & \multirow{2}{*}{measure} & \multicolumn{3}{c|}{sd=0.05} & \multicolumn{3}{c|}{sd=0.10} & \multicolumn{3}{c}{sd=0.15} \\ \cline{3-11} 
 &  & $R_{sm}$ & $R_{pd}$ & $l$ or $v$ & $R_{sm}$ & $R_{pd}$ & $l$ or $v$ & $R_{sm}$ & $R_{pd}$ & $l$ or $v$ \\ \hline
\multirow{2}{*}{hierarchical} & silhouette & 32.94 & 32.94 & 1.3 & 10.06 & 10.06 & 2.7 & 2.85 & 2.85 & 4.0 \\
 & dissimilarity & 32.86 & 32.86 & 2.1 & 9.86 & 9.86 & 4.3 & 2.44 & 2.44 & 6.6 \\ \hline
\multirow{2}{*}{k-means} & silhouette & 33.78 & 33.67 & 1.4 & 12.96 & 12.71 & 2.7 & 4.98 & 4.60 & 4.2 \\
 & dissimilarity & 34.13 & 34.03 & 2.1 & 12.87 & 12.66 & 4.3 & 5.26 & 4.94 & 6.6
\end{tabular}
\caption{For the three-group continuous datasets, the soft-misclassification rates $R_{sm}$ and partition-disagreement rates $R_{pd}$ as percents for different values of ${\rm sd}(P_{h1})$ and two different clustering methods. The value of the tuning parameter, $l$ or $v$, to achieve the ${\rm sd}(P_{h1})$ is also shown} 
\label{tab:tuning_cont_3clst}
\end{table}
\end{subtables}

The performance of the two measures depends on the clustering method. Referring to Table \ref{tab:tuning_cont_3clst}, k-means clustering has slightly higher soft-misclassification and partition-disagreement rates than hierarchical clustering for both measures, because k-means clustering is less stable and occasionally generates unreasonable clustering solutions, depending on the starting point chosen for the algorithm.

\subsection{Comparison to Other Clustering Methods}

\subsubsection{Binary Data}

As benchmarks, we compute the estimated posterior probabilities in the sample for the FANNY algorithm, LCA and Gaussian model-based clustering. Table \ref{tab:comparison_binary} summarizes results for all the clustering methods for the binary data.
In the FANNY algorithm, the user-specified, exponent parameter, $r$, affects the soft-misclassification rate and partition-disagreement rate for the data. In contrast to the tuning parameters $l$ and $v$ of the proposed measures, increasing the FANNY parameter, $r$, creates fuzzier clusters and leads to more uncertainty for both non-hybrid and hybrid individuals. Increasing, $r$, leads to an increase in the overall $R_{sm}$ and $R_{pd}$, just like decreasing, $v$ and $l$, of the proposed measures. Therefore, FANNY's tuning parameter, $r$, also represents a tradeoff between detecting hybrid individuals and misclassifying non-hybrid individuals. In Table \ref{tab:comparison_binary}, these tuning parameters are set to default values of $v=1$, $l=1$ and $r=2$. FANNY does well at detecting the hybrid individuals, whose measures of cluster-membership certainty appear to be normally distributed with a mean around 0.50 for the two-group datasets or 0.33 for the three-group datasets (results not shown). However, for the SMD matrix the larger values of FANNY's soft-misclassification and partition-disagreement rates suggest that decreasing $r$ may be necessary, even though this will increase the variance of the hybrid's measure of cluster certainty. Although the SMD matrix is good enough for PAM to identify the three clusters, it has relatively fuzzy similarities and thus shows large $R_{sm}$ and $R_{pd}$ with the default value of the tuning parameters. The {\tt PReMiuM} co-occurrence matrix hass easily distinguishable similarities between individuals and thus shows small $R_{sm}$ and $R_{pd}$. 

For LCA and Gaussian model-based clustering, we observe extreme behavior for the hybrid's cluster-membership certainties: the hybrid individual always has an estimated posterior probability of either zero or one (results not shown), with equiprobable assignment to either extreme. These estimated posterior probabilities are incompatible with how the hybrid data were simulated. Although all the posterior probability estimators appear to be unbiased (the point estimates are within simulation error of 0.50 or 0.33, with 95\% confidence), the standard deviations are very close to the maximum of 0.50 for two-group datasets or 0.47 for three-group datasets. That is, a hybrid individual is randomly assigned to one of the clusters with an estimated posterior probability equal to one. 
   
\begin{table}[htbp]
\centering
\begin{tabular}{cc|cccc|cccc}
 &  & \multicolumn{4}{c}{two-group} & \multicolumn{4}{|c}{three-group} \\ \cline{3-10}
 &  & mean & sd & $R_{sm}$ & $R_{pd}$ & mean & sd & $R_{sm}$ & $R_{pd}$ \\ \hline
\multicolumn{1}{c|}{\multirow{3}{*}{\begin{tabular}[c]{@{}c@{}}silhouette\\ ($l$=1)\end{tabular}}} & Euclidean & 0.50 & 0.16 & 12.66 & 12.66 & 0.34 & 0.15 & 23.77 & 23.75 \\
\multicolumn{1}{c|}{} & SMD & 0.50 & 0.07 & 28.24 & 28.24 & 0.34 & 0.03 & 31.83 & 31.99 \\
\multicolumn{1}{c|}{} & {\tt PReMiuM} & 0.50 & 0.27 & 2.12 & 2.12 & 0.34 & 0.27 & 4.37 & 4.28 \\ \hline
\multicolumn{1}{c|}{\multirow{3}{*}{\begin{tabular}[c]{@{}c@{}}dissimilarity\\ ($v$=1)\end{tabular}}} & Euclidean & 0.50 & 0.11 & 19.81 & 19.81 & 0.34 & 0.11 & 34.21 & 34.19 \\
\multicolumn{1}{c|}{} & SMD & 0.50 & 0.04 & 35.87 & 35.87 & 0.33 & 0.02 & 58.93 & 58.94 \\
\multicolumn{1}{c|}{} & {\tt PReMiuM} & 0.50 & 0.23 & 3.72 & 3.72 & 0.34 & 0.23 & 7.31 & 7.24 \\ \hline
\multicolumn{1}{c|}{\multirow{3}{*}{\begin{tabular}[c]{@{}c@{}}FANNY\\ ($v$=2)\end{tabular}}} & Euclidean & 0.50 & 0.14 & 11.60 & 11.60 & 0.34 & 0.14 & 22.06 & 22.04 \\
\multicolumn{1}{c|}{} & SMD & 0.50 & 0.05 & 32.21 & 32.21 & 0.33 & 0.00 & 66.67 & 66.67 \\
\multicolumn{1}{c|}{} & {\tt PReMiuM} & 0.50 & 0.24 & 1.13 & 1.13 & 0.34 & 0.24 & 2.47 & 2.67  \\ \hline
\multicolumn{1}{c|}{\multirow{2}{*}{model-based}} & LCA & 0.52 & 0.50 & 0.00 & 0.00 & 0.32 & 0.46 & 0.23 & 0.02 \\
\multicolumn{1}{c|}{} & Gaussian & 0.52 & 0.47 & 0.00 & 0.00 & 0.33 & 0.43 & 0.28 & 0.13
\end{tabular}
\caption{Comparison of the results for all the clustering methods for the binary data. The soft-misclassification rate, $R_{sm}$ and the partition-disagreement rate, $R_{pd}$ are shown in percents}
\label{tab:comparison_binary}
\end{table}

\subsubsection{Continuous Data}

For the continuous datasets, we compute the estimated posterior probabilities for the FANNY algorithm and Gaussian model-based clustering. Table \ref{tab:comparison_cont} summarizes results for all the clustering methods, where the tuning parameters are set to default values of $v=1$, $l=1$ and $r=2$. The soft-misclassification and partition-disagreement rates for FANNY are close to 50.00\% for the two-group datasets and 66.67\% for the three-group datasets are what we expect from a random assignment. Large rates indicate that more tuning is necessary to generate crisper clusters.

All the methods generate unbiased estimators; i.e., 0.50 for the two-group datasets and 0.33 for the three-group datasets. However, the hybrids' estimated posterior probabilities under the Gaussian mixture model are extreme (i.e., either 0 or 1; results not shown), which indicates that the Gaussian mixture model fails to detect the hybrid individuals.

\begin{table}[htbp]
\centering
\begin{tabular}{cc|cccc|cccc}
                                                   &                       & \multicolumn{4}{c}{two-group}       & \multicolumn{4}{|c}{three-group}    \\ \cline{3-10} 
                                                   &                       & mean & sd   & $R_{sm}$ & $R_{pd}$ & mean & sd   & $R_{sm}$ & $R_{pd}$ \\ \hline
\multicolumn{1}{c|}{\multirow{2}{*}{hierarchical}} & silhouette ($l=1$)    & 0.50 & 0.07 & 21.60     & 21.60     & 0.33 & 0.03 & 40.43     & 40.43     \\
\multicolumn{1}{c|}{}                              & dissimilarity ($v$=1) & 0.50 & 0.04 & 30.11     & 30.14     & 0.33 & 0.04 & 40.40     & 40.40     \\ \hline
\multicolumn{1}{c|}{\multirow{2}{*}{k-means}}      & silhouette ($l=1$)    & 0.50 & 0.07 & 21.59     & 21.59     & 0.33 & 0.03 & 42.47     & 42.37     \\
\multicolumn{1}{c|}{}                              & dissimilarity ($v$=1) & 0.50 & 0.04 & 30.10     & 30.10     & 0.33 & 0.04 & 41.93     & 41.85     \\ \hline
\multicolumn{2}{c|}{FANNY ($r$=2)}                                         & 0.50 & 0.06 & 45.18     & 45.18     & 0.33 & 0.01 & 65.04     & 65.04     \\ \hline
\multicolumn{2}{c|}{Gaussian model-based}                                  & 0.48 & 0.50 & 0.00      & 0.00      & 0.32 & 0.46 & 0.00      & 0.00     
\end{tabular}
\caption{Comparison of the results for all the clustering methods for the continuous data. The soft-misclassification rate, $R_{sm}$ and the partition-disagreement rate, $R_{pd}$ are shown in percents}
\label{tab:comparison_cont}
\end{table}

\section{A Real Data Example}

We use the famous Fisher's iris data set (see \citealp{Fisher1936}) as an example to assess the performance of the proposed measures. Fisher's iris data is considered as a benchmark for clustering methods and has attracted much work in statistical analysis. This data set consists of measurements of sepal length and width, and petal length and width, in centimeters for 150 irises. There are three species of iris: \emph{setosa}, \emph{versicolor} and \emph{virginica}, each consisting of 50 individuals. Fisher found that although the \emph{setosa} species can be neatly separated, the other two groups, \emph{versicolor} and \emph{virginica} are difficult to distinguish. 

\begin{figure}[htbp]
\centering
\includegraphics[width=2.8in]{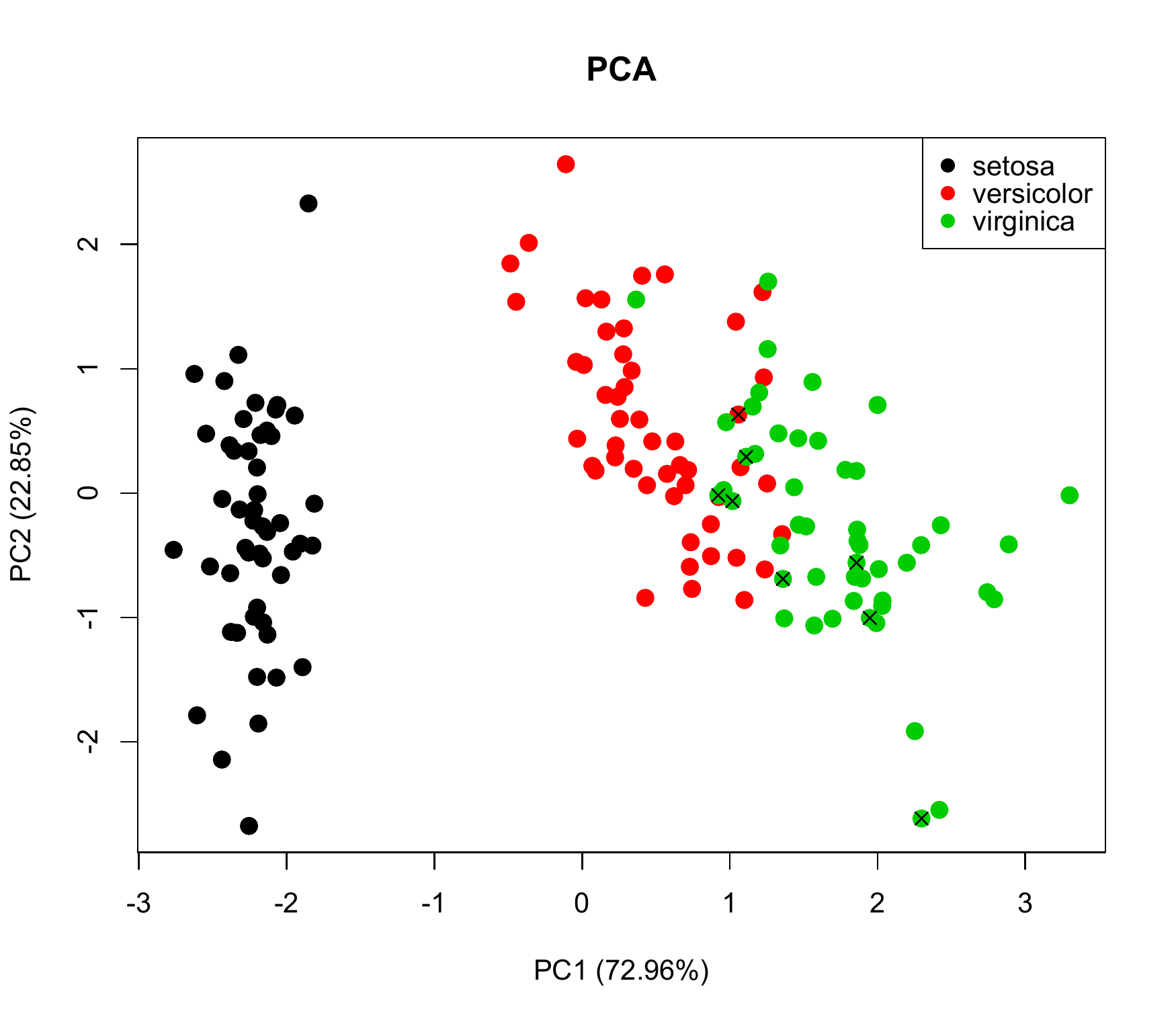}
\caption{A plot based on the first two principal components for the Fisher's iris data. The \emph{setosa} (black) is neatly separated, and the \emph{versicolor} (red) and \emph{virginica} (green) are difficult to distinguish. The individuals misclassified by hierarchical clustering are marked by ``$\times$''.}
\label{fig:PCA_iris}
\end{figure}

We use Orl\'oci's chord distance as the dissimilarity measure and hierarchical clustering for the iris data. The chord distance (see \citealp{Orloci1967}) between two individuals $\mathbf{x}$ and $\mathbf{y}$ is defined as
\begin{equation*}
D_{chord} = \sqrt{\sum_{i=1}^p (\frac{x_{i}}{\sqrt{\sum_{j=1}^p x_j^2}}-\frac{y_i}{\sqrt{\sum_{j=1}^p y_j^2}})^2},
\end{equation*}
where $p$ is the number of features. The chord distance is the Euclidean distance computed after scaling the feature vectors to length 1 and thus can solve problems caused by differing scales of measurement. Hierarchical clustering is based on the chord distance matrix and the clustering result is shown in Table \ref{tab:iris_clustering}. Hierarchical clustering manages to neatly separate \emph{setosa} but fails to completely distinguish \emph{versicolor} and \emph{virginica}. Figure \ref{fig:PCA_iris} is a plot based on the first two principal components, with black, red and green representing \emph{setosa}, \emph{versicolor} and \emph{virginica}, respectively. The individuals misclassified by hierarchical clustering are denoted by ``$\times$''.

\begin{table}[ht]
\centering
\begin{tabular}{c|ccc}
cluster & setosa & versicolor & virginica \\ \hline
1       & 50     & 0          & 0         \\
2       & 0      & 49         & 7         \\
3       & 0      & 1          & 43       
\end{tabular}
\caption{The clustering results of hierarchical clustering for Fisher's iris data}
\label{tab:iris_clustering}
\end{table}

\begin{figure}
\centering
\includegraphics[height=3in, width=5in]{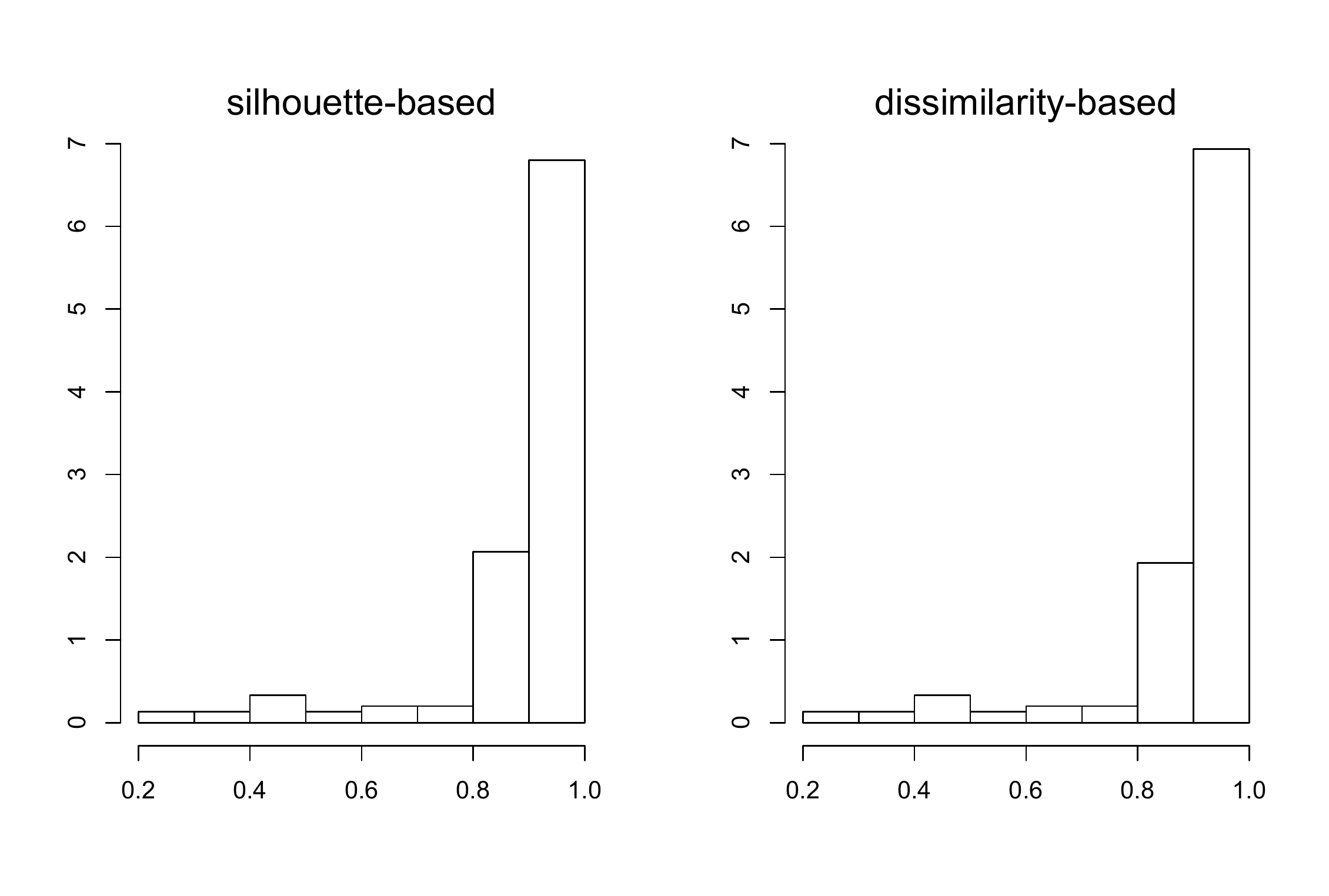}
\caption{The distribution of the individual cluster-membership certainties for the assigned cluster; silhouette-based measure (left) and dissimilarity-based measure (right)}
\label{fig:HistVague}
\end{figure}

We tune the parameters $l$ and $v$ so that the soft-misclassification rate of the proposed measures is 10\%. Referring to equation \eqref{eq:softmis}, the average certainty measure, $\frac{1}{N} \sum_{i=1}^N P_{ig_i}^q$, over all individuals is thus $100\%-10\%=90\%$. In this case, the rates of partition-disagreement and soft-misclassification are very similar, to within $\pm$ 0.8\% (results not shown). The individual cluster-membership certainty of the misclassified individuals is shown in Table \ref{tab:mis_certainty}. All of the eight misclassified individuals' certainties are less than 0.90, for both the true group and the assigned cluster. More importantly, six misclassified individuals have a higher cluster-membership certainty of the true group (emboldened) than of the assigned cluster (in italics). For example, the silhouette-based measures of individual 111 belonging to \emph{setosa}, \emph{versicolor} and \emph{virginica} are 0.01, 0.33 and 0.66, respectively. We can infer that individual 111 is very unlikely to belong to the \emph{setosa} group and has a higher probability of fitting in the \emph{virginica} than the \emph{versicolor} group. Though the hierarchical clustering method incorrectly assigns it to the \emph{versicolor} group, the silhouette-based measure manages to detect the discrepancy. In this way, our proposed measure may help users identify potentially misclassified individuals in the original partition.
Figure \ref{fig:HistVague} shows the distribution of the individuals' cluster-membership certainty for their assigned clusters. For both measures, most of the cluster-membership certainties are greater than 0.80, indicating a generally good cluster-membership assignment. The 5\% sample quantiles for the silhouette and dissimilarity-based measures, which are 0.48 for both measures, are used as the threshold for identifying ambiguous individuals. Eight individuals which fall below this threshold are on the edge of their true group and neighbor group. Six of them are shown in Table \ref{tab:mis_certainty}; they are individuals 111, 126, 128, 130, 134 and 139. However, two individuals that are correctly classified by the clustering method fall below this threshold, too; they are reported in Table \ref{tab:vague_certainty}. 

\begin{table}[htbp]
\centering
\begin{tabular}{c|ccc|ccc}
\multicolumn{1}{l|}{} & \multicolumn{3}{c|}{silhouette} & \multicolumn{3}{c}{dissimilarity} \\ \hline
index & \emph{setosa}   & \emph{versicolor}   & \emph{virginica}  & \emph{setosa}   & \emph{versicolor}   & \emph{virginica}   \\  \hline
84     & 0.00     & \textbf{0.13}         & \textit{0.87}       & 0.00     & \textbf{0.12}         & \textit{0.88}       \\
111    & 0.01     & \textit{0.33}         & \textbf{0.66}       & 0.00     & \textit{0.35}         & \textbf{0.65}       \\
126   & 0.00     & \textit{0.22}         & \textbf{0.78}       & 0.00     & \textit{0.23}         & \textbf{0.77}       \\
128   & 0.00     & \textit{0.28}         & \textbf{0.71}       & 0.00     & \textit{0.30}         & \textbf{0.70}       \\
130   & 0.01     & \textit{0.46}         & \textbf{0.53}       & 0.00     & \textit{0.47}         & \textbf{0.53}       \\
132   & 0.01     & \textit{0.52}         & \textbf{0.48}       & 0.00     & \textit{0.52}         & \textbf{0.48}       \\
134   & 0.01     & \textit{0.45}         & \textbf{0.55}      & 0.00     & \textit{0.46}         & \textbf{0.54}       \\
139   & 0.01     & \textit{0.32}         & \textbf{0.68}       & 0.00     & \textit{0.33}         & \textbf{0.67}            
\end{tabular}
\caption{The individual cluster-membership certainty for the misclassified individuals under hierarchical clustering. The true membership for each individual is emboldened and the assigned cluster is in italics.}
\label{tab:mis_certainty}
\end{table}

\begin{table}[htbp]
\centering
\begin{tabular}{c|ccc|ccc}
    & \multicolumn{3}{c|}{silhouette}    & \multicolumn{3}{c}{dissimilarity} \\ \cline{2-7} 
    & \emph{setosa} & \emph{versicolor}     & \emph{virginica}      & \emph{setosa} & \emph{versicolor}     & \emph{virginia}       \\ \hline

71  & 0.01   & \textbf{0.40} & 0.59          & 0.00   & \textbf{0.42} & 0.58          \\
73  & 0.01   & \textbf{0.47} & 0.53          & 0.00   & \textbf{0.47} & 0.53      
\end{tabular}
\caption{The 5\% sample quantiles for the silhouette- and dissimilarity-based uncertainty measures of the assigned clusters are both 0.48. Individuals whose cluster-membership certainty of the assigned cluster is below the 5\% sample quantiles and not shown in Table 6 are presented. The true groups, which are also the assigned clusters, are emboldened.}
\label{tab:vague_certainty}
\end{table}

\section{Discussion}

Pairwise dissimilarities between individuals reflect the structure present in multivariate data and provide key information for clustering individuals. The silhouette value uses pairwise dissimilarities to measure how well an individual fits to the cluster it has been assigned relative to the other clusters in a hard partition. We have proposed two probability-like measures of cluster-membership certainty, one which extends the classic silhouette and the other based directly on the dissimilarities. These measures can assist with identifying individuals of ambiguous cluster membership after applying a hard-clustering algorithm. As they are simple and behave like probabilities, they may be conveniently applied in clinical and bioinformatics settings which use hard partitional clustering to explore data (e.g., \citealp{Molitor2010} and \citealp{Liverani2015}). 

Being model-free, the two proposed measures highly depend on the quality of the dissimilarity matrix. A dissimilarity matrix that highlights the relevant property of the features and captures the data structure is expected to enhance the performance of the proposed measures. It is important to note that the maximum of each individual's cluster-membership certainty may not necessarily correspond to the assigned cluster, as seen in the iris data analysis. In the iris data, this discrepancy occurred when the individuals were on the border of two neighboring clusters. Such discrepancies can help users identify the possible misclassified individuals. 

In our simulations, both proposed measures and all the soft-clustering methods provide unbiased measures of the hybrid individual's probability of cluster membership. However, the measures from the model-based methods have a U-shaped distribution and high variance. In contrast, our measures and those from the FANNY algorithm can be tuned to have a distribution with a mode at 0.5 for two-group datasets or 0.33 for three-group datasets. For example, the application of PAM/k-means/hierarchical clustering and Gaussian model-based clustering to the Euclidean distance matrix contrasts the behavior of the proposed measures to the model-based clustering, in this regard. Our measures are able to estimate  the hybrid's ambiguous membership whereas the model-based clustering methods are not.

One feature that our proposed measures and the FANNY algorithm have in common is the way the exponent parameter changes the variance of the hybrid's cluster membership, the soft-misclassification rate and partition-disagreement rate in the sample. In general, increasing the exponent parameter leads to an increased variance for the hybrid's estimated cluster-membership certainty and a decrease of soft-misclassification rate and partition-disagreement rate for the rest of the sample. The tuning parameters $l$ and $v$ represent a tradeoff between the soft-misclassification rate and the partition-disagreement rate for non-hybrids on the one hand and the variance of the hybrid's cluster-membership measure on the other. We recommend that researchers experiment with tuning $l$ or $v$ to balance this tradeoff. A similar recommendation has been proposed for the tuning parameter $r$ in the FANNY algorithm (see \citealp{Kaufman1990}). For the proposed measures, we suggest that a soft-misclassification rate around 10\% is a sign of good cluster membership, as the individuals are assigned to their clusters with an average certainty of 90\%. When the true assignment is unknown, i.e., the soft-misclassification rate cannot be computed, researchers can use the partition-disagreement rate instead. The sample distribution of cluster-membership certainties for each individual's assigned cluster can assist users to determine a threshold of ambiguous certainty. For example, we have used the 5\% quantiles in our illustration with the iris data.

In our simulations, the proposed measures reflect the hybrid individual's probability of cluster membership as expected, though their soft-misclassification rates and partition-disagreement rates are higher than FANNY's. The higher soft-misclassification rates and partition-disagreement rates are expected since the proposed measures work from a fixed clustering while FANNY has more flexibility to simultaneously cluster and assign fuzzy memberships.

In summary, our measures are straightforward to implement and worth considering as a way to augment hard-clustering methods which give no measure of the posterior probabilities of cluster membership for individuals. Determining the number of clusters, however, is beyond the scope of this work.









\bibliographystyle{plainnat}

\bibliography{Reference}
\end{document}